 \documentclass[final,5p,times,twocolumn]{elsarticle}

\usepackage{amssymb}
\usepackage{lipsum}
\usepackage{graphicx}
\usepackage{bm}
\usepackage{tikz}
\usepackage{pgfplots}
\pgfplotsset{compat=newest}
\usepackage[colorlinks,citecolor=blue,urlcolor=blue,linkcolor=blue]{hyperref}
\usepackage{amsmath, physics}

\usepackage{orcidlink}
\journal{Physics Letters B}

\begin{document}

\begin{frontmatter}



%\title{Schwarzschild-de Sitter thermodynamics and logarithmic correction of the entropy}
\title{Finite Hilbert space and maximum mass of Schwarzschild black holes from a Generalized Uncertainty Principle}

\author[1,2,3]{S. Jalalzadeh } 
\affiliation[1]{organization={Izmir Institute of Technology},%Department and Organization
            addressline={Department of Physics}, 
            city={Urla},
            postcode={35430}, 
            state={Izmir},
            country={Türkiye}}
\affiliation[2]{organization={Dogus University},%Department and Organization
            addressline={Department of Physics}, 
            city={Dudullu-Ümraniye},
            postcode={34775}, 
            state={Istanbul},
            country={Türkiye}}
\affiliation[3]{organization={Khazar University},%Department and Organization
            addressline={Center for Theoretical Physics}, 
            city={41 Mahsati Str.},
            postcode={AZ1096}, 
            state={Baku},
            country={Azerbaijan}}               
\ead[2]{shahramjalalzadeh@iyte.edu.tr}          
\author[4]{H. Moradpour}
\affiliation[4]{organization={Research Institute for Astronomy and Astrophysics of Maragha (RIAAM)},%Department and Organization
            addressline={University of Maragheh}, 
            city={Maragheh},
            postcode={55136-553}, 
          %  state={PE},
            country={Iran}}            
\ead[1]{h.moradpour@riaam.ac.ir}

%%%%%%%%%%%%%%%%%%%%%%%%%%%%%%%%%%%%%%%%%%%%%%%%%%%%%%%%%%%%%%%%%%%%%%%%%%%%%%%%%%%%%%%%%%%%%%%%%%%%%%%%%%%%%%%%%
\begin{abstract}
We show that implementing a generalized uncertainty principle (GUP) with both minimal length and maximal momentum directly on the reduced phase space of the Schwarzschild black hole (BH) leads to a finite and discrete mass spectrum, a strict upper bound on the BH mass, a bounded entropy, and a fully regulated Hawking temperature.  
We further construct a GUP-deformed lapse function that preserves the ADM mass and horizon radius while exactly reproducing the GUP temperature through the surface gravity.  Using the most massive observed supermassive BHs, we derive the constraint on the GUP parameter, $\beta\lesssim 10^{-98}$, showing that present astrophysical data already impose robust bounds on minimal length quantum gravity.
\end{abstract}




\end{frontmatter}

%\tableofcontents


\section{Introduction}\label{Sec1}

Many approaches to quantum gravity, including string theory, loop quantum gravity, noncommutative geometry, and deformed special relativity, indicate the existence of a fundamental minimal length \cite{Hossenfelder:2012jw}. A common phenomenological realization is a GUP, which modifies the Heisenberg algebra and leads to a minimal position uncertainty and maximal momentum \cite{KMM1995, Scardigli1999, Ali:2009zq}. BHs provide a natural arena to probe such Planck scale effects, and numerous studies have explored GUP corrections to Hawking radiation, remnants, and thermodynamics \cite{Adler:2001vs, Bina:2010ir, Chen:2022dap, Gine:2025qsb, Gingrich:2024mgk, Lambiase:2022xde, Casadio:2020ueb, Vagenas:2019rai, Alonso-Serrano:2018ycq}.

A fundamental minimal length suggests that the phase space structure of gravity must be modified at high energies. The Schwarzschild geometry is exceptionally suited for this purpose because its Hamiltonian reduction leads to a single canonical pair, which can be unwrapped into an equivalent harmonic oscillator phase space. When a higher-order GUP with maximal momentum is implemented in this reduced description, the allowed phase space volume becomes compact, and the associated oscillator admits only a finite number of quantum states. As a result, the permitted BH masses form a discrete set that terminates at a maximum value fixed by the GUP parameter. This mechanism establishes a transparent connection between minimal length physics and a finite BH spectrum and implies that the number of accessible BH microstates is intrinsically finite.

The presence of a maximal momentum in the GUP algebra plays an essential role. It not only induces the upper mass bound but also regulates the thermodynamic behavior of the BH. Earlier studies introduced GUP corrections to either the spectrum, the Hawking temperature, or the entropy, yet these analyses were typically heuristic or incomplete. They did not provide a single framework in which the same GUP simultaneously determines the mass spectrum, the thermodynamic quantities, and the deformation of the spacetime metric. By contrast, the reduced phase space quantization integrates all these elements into a consistent scheme in which the discrete spectrum, the finite entropy, the regulated temperature, and the metric deformation originate from the same microscopic input. This unified structure gives the BH a finite-dimensional Hilbert space and brings GUP phenomenology into direct contact with gravitational geometry.

However, a fully self-consistent framework in which the GUP is implemented directly on the reduced phase space of the Schwarzschild BH, yielding simultaneously a finite mass spectrum, bounded entropy, regulated temperature, and a geometrically consistent metric deformation, has remained incomplete. Earlier work focused on heuristic mass quantization, effective temperature corrections, or specific remnants, but not on a unified canonical construction.

In this article, we fill this gap. Building on the Hamiltonian reduction of the Schwarzschild geometry \cite{Kuchar:1994zk, Louko:1996md, Das:2001ic, Jalalzadeh:2026ttf, Jalalzadeh:2026ssq}, we map the BH to a $1D$ harmonic oscillator in the $(x,p)$ plane and quantize it using a higher-order GUP that includes both minimal length and maximal momentum \cite{Pedram:2012}. This yields (i) a finite and discrete mass spectrum $M_n$, (ii) a strict upper bound on the BH mass $M_{\max}^2 < m_P^2/(2\beta)$, (iii) a bounded entropy $S_{\max}<\pi/\beta$, and (iv) a GUP-modified Hawking temperature that remains finite at the endpoint of the spectrum. We then reconstruct a GUP-deformed metric that preserves the ADM mass and horizon location while reproducing exactly the GUP temperature from the surface gravity, and we show that its Newtonian limit contains only short-range corrections compatible with observations. Finally, using the largest known supermassive BH masses, we extract an astrophysical upper bound on the GUP parameter.

It is useful to contrast the present framework with earlier attempts to incorporate GUP effects into BH physics. The closest analysis in structure is the work of Bina, Jalalzadeh, and Moslehi \cite{Bina:2010ir}, which applied a GUP to a $1D$ Schrödinger problem and obtained corrections to the BH mass spectrum and Hawking temperature. However, their treatment did not arise from a canonical reduction of the Schwarzschild geometry, and as a result, the spectrum, entropy, and thermodynamics were not tied to the underlying gravitational phase space.

Pedram's study of the same higher-order GUP algebra \cite{Pedram:2012} established the existence of a maximal momentum and a compact phase space for a quantum particle, providing exact solutions to the corresponding GUP-modified Schrödinger equation. Nevertheless, the analysis was purely quantum mechanical and did not address gravitational systems. In particular, the construction did not specify how the GUP should modify the canonical variables of a BH, nor how the resulting quantum spectrum would feed back into the geometry. The present work builds directly on Pedram's algebra. Still, it embeds it into the reduced Schwarzschild phase space, thereby linking the algebraic structure to BH mass quantization, entropy, and geometry in a single consistent framework.

Kunstatter's early result \cite{Kunstatter:2002pj} showed that the reduced phase space of a spherically symmetric BH admits an oscillator-like adiabatic invariant, which in turn leads to an approximately evenly spaced area spectrum. This analysis provided the conceptual foundation for treating the BH as a $1D$ system, but it did not include any form of GUP or minimal length modification. Consequently, the spectrum was unbounded, the entropy remained monotonic, and no maximal mass or finite-state structure emerged. In contrast, we show that introducing a higher-order GUP directly into the same reduced phase space yields a compactification of the allowed phase-space volume and produces a finite sequence of admissible BH states.

Finally, the seminal work of Adler, Chen, and Santiago \cite{Adler:2001vs} argued that GUP effects may halt Hawking evaporation and lead to BH remnants. Their analysis focused primarily on the infrared behavior of the Hawking temperature and the possible existence of a minimal mass, but it did not predict a maximal mass or a finite number of states. Nor did it provide a deformation of the Schwarzschild lapse compatible with the modified temperature. The present approach reproduces the regularization of the temperature at small mass while also predicting a strict upper bound on the BH mass and a finite Hilbert space, both absent in earlier remnant-based models.

Taken together, these comparisons show that previous analyses captured isolated aspects of GUP physics, such as modified spectra or modified temperatures, but none provided a unified picture in which the GUP simultaneously determines the BH mass spectrum, entropy, thermodynamics, and spacetime geometry. The reduced phase space formulation employed here makes this unification possible by placing all quantum and geometric quantities under the same canonical structure.

The remainder of this paper is organized as follows. In Sec. \ref{Sec2} we recall the reduced phase space of the Schwarzschild BH and implement the higher-order GUP directly in the canonical variables, obtaining a finite and discrete mass spectrum. In Sec. \ref{Sec3} we obtained the associated entropy, temperature, and specific heat of the BH. In Sec. \ref{Sec3} we use the GUP-regulated temperature to reconstruct a consistent deformation of the Schwarzschild lapse that preserves the ADM mass and horizon radius while encoding minimal-length effects in the near-horizon geometry. Section \ref{Sec4} analyzes the Newtonian limit, observational implications, and the resulting bound on the GUP parameter from the most massive SMBHs. We conclude in Sec. \ref{Sec5} with a discussion of the physical significance of a finite BH Hilbert space and the broader implications for quantum gravity.

%%%%%%%%%%%%%%%%%%%%%%%%%%%%%%%%%%%%%%%%%%%%%%%%%%%%%%%%%%%%%%%%%%%%%%%%%%

\section{GUP quantization in Reduced phase space}\label{Sec2}
The spherically symmetric ADM line element can be written as \cite{Kuchar:1994zk}
\begin{equation}
\mathrm ds^{2} = -N^{2}\mathrm dt^{2} + \Lambda^{2}(\mathrm dr+N^{r} \mathrm dt)^{2} + R^{2}\mathrm d\Omega^{2},
 \label{eq:ADM}
\end{equation}
with canonical pairs $(\Lambda,P_\Lambda)$ and $(R,P_R)$.
After solving the Hamiltonian and momentum constraints and performing the canonical transformation constructed in Ref.~\cite{Kuchar:1994zk, Louko:1996md, Jalalzadeh:2021gtq}, the Einstein--Hilbert action functional reduces to the following action,
\begin{equation}
 S_{\rm red} = \int \left(P_M \dot M - M\right)\,\mathrm dt,
 \qquad \{M,P_M\}=1,
 \label{eq:Sred}
\end{equation}
so the only gauge-invariant degree of freedom is the ADM mass $M$ with conjugate momentum $P_M$.

The conjugate momentum of ADM mass represents the asymptotic time coordinate at the spacelike slice. Since $P_M$ plays the role of time, it should be periodic \cite{Louko:1996md, Jalalzadeh:2022rxx}, with the period being the inverse Hawking temperature. Therefore,
\begin{equation}
 P_M \sim P_M + \frac{1}{T_H},
 \qquad
 T_H = \frac{m_P^2}{8\pi M},
\end{equation}
where $T_H$ is the Hawking temperature, and $m_P=1/\sqrt{G}$ is the Planck mass in natural units ($\hbar=c=k_B=1$). The periodicity of \( P_M \) guarantees there is no conical singularity in the \( 2D \) Euclidean section near the BH horizon \cite{Das:2001ic}. However, this identification suggests that the physical phase space is a wedge taken from the entire $(M, P_M)$ plane, restricted by the $M$ axis and the line $P_M = 1/T_H(M)$.  
The physical phase space is thus a wedge in $(M, P_M)$.
As shown in \cite{Das:2001ic, Medved:1998ks}, this wedge can be unwrapped into the full plane via the canonical map,
\begin{equation}
\begin{split}
 x=\sqrt{\frac{A}{4\pi G}}\cos(2\pi P_M T_H),~~~
 p=\sqrt{\frac{A}{4\pi G}}\sin(2\pi P_M T_H),
 \end{split}
\end{equation}
where $A=16\pi G^2 M^2$ is the surface area of the event horizon. Squaring and adding the above gives
\begin{equation}
 x^2 + p^2 = \frac{A}{4\pi G} = 4GM^2 = \frac{4M^2}{m_P^2}.
 \label{eq:x2p2}
\end{equation}
The reduced Schwarzschild dynamics is therefore equivalent to those of a unit-frequency harmonic oscillator with
%\begin{equation}
 $E_{\rm osc} = \frac{1}{2}(x^2+p^2) = \frac{2M^2}{m_P^2}$,
 %\label{eq:Eosc}
%\end{equation}
providing a canonical origin for equally spaced spectra in $M^2$ (and area) in the non-GUP case.

%%%%%%%%%%%%%%%%%%%%%%%%%%%%%%%%%%%%%%%%%%%%%%%%%%%%%%%%%%%%%%%%%%%%%%%
\subsection{GUP quantization and finite mass spectrum}
We now promote $(x,p)$ to operators satisfying Pedram's higher-order GUP algebra \cite{Pedram:2012}
\begin{equation}
 [\hat x,\hat p] = \frac{i}{1-\beta \hat p^2}, \qquad \beta>0,
 \label{eq:GUP}
\end{equation}
which implies both a minimal length and a maximal momentum $|p|\le 1/\sqrt{\beta}$, and reproduces the usual GUP structure at low momenta. Note that the above commutator relation agrees with KMM's (Kempf, Mangano, and Mann) \cite{Kempf:1994su} and Nouicer’s GUP \cite{Nouicer:2007jg} to the leading orders. Since in our work $x$ and $p$ are dimensionless quantities, the GUP parameter $\beta$ is also dimensionless.

In the momentum representation, the quantization map is given by \cite{Pedram:2012}
\begin{equation}
 \hat p\,\psi(p)=p\,\psi(p),\quad
 \hat x\,\psi(p)=\frac{i}{1-\beta p^2}\frac{\mathrm d\psi}{\mathrm dp},
 \quad |p|\le\frac{1}{\sqrt{\beta}}.
 \label{eq:rep}
\end{equation}
Then, the constraint equation \eqref{eq:x2p2} yields the following $1D$ Wheeler--DeWitt equation:
\begin{multline}
 \frac{\mathrm d^2\psi}{\mathrm dp^2}
 + \frac{2\beta p}{1-\beta p^2}\frac{\mathrm d\psi}{\mathrm dp} 
 + (1-\beta p^2)^2\left(\frac{M^2}{m_P^2}-p^2\right)\psi(p)=0.
 \label{eq:SchrGUP}
\end{multline}

As we are interested in the thermodynamics of massive BHs, we use the semiclassical method to obtain the mass spectrum of the above equation. Bohr--Sommerfeld quantization, adapted to the GUP measure \cite{Pedram:2012}, leads to the discrete mass spectrum.
%%%%%%%%%%%%%%%%%%%%%%%%%%%%%%%%%%%
The semiclassical quantization condition follows from the action variable
\begin{equation}
I=\oint x\,\mathrm dp = 2\pi\!\left(n+\tfrac12\right),
\end{equation}
where $x(p)=\sqrt{4M^{2}/m_{\rm P}^{2}-p^{2}}$ and the GUP modifies the integration measure through the Jacobian $(1-\beta p^{2})^{-1}$:
\begin{equation}
2\int_{0}^{p_*}\!\frac{\mathrm dp}{1-\beta p^{2}}
\sqrt{\frac{4M^{2}}{m_{\rm P}^{2}}-p^{2}}
=
2\pi\!\left(n+\tfrac12\right),
\end{equation}
with $p_{*}=\min\!\left(\tfrac{2M}{m_{\rm P}},1/\sqrt{\beta}\right)$.  
Evaluating the integral yields
\begin{equation}
 M_{n}^{2}
 = \frac{m_{\rm P}^{2}}{2\beta}
   \left[1 - \sqrt{1 - 2\beta\!\left(n+\tfrac12\right)}\right],
 \quad n=0,1,\dots,n_{\max},
 \label{eq:Mn2}
\end{equation}
with a finite number of levels,
\begin{equation}
 n_{\max}
 = \left\lfloor \frac{1}{2\beta}-\frac12\right\rfloor,
 \label{eq:nmax}
\end{equation}
where $\lfloor x\rfloor$ is the floor function, $\lfloor x\rfloor=\max\{m\in \mathbb Z|m\leq x\}$.

For $\beta\ll 1$, expanding the square root gives,
\begin{equation}
 M_{n}^{2}\simeq
 \frac{m_{\rm P}^{2}}{2}\Bigl(n+\tfrac12\Bigr)
 \left[1 - \frac{\beta}{2}\Bigl(n+\tfrac12\Bigr)\right]
 + \mathcal{O}(\beta^{2}),
 \label{eq:Mn-exp}
\end{equation}
i.e., an equispaced $M^{2}$ spectrum plus a small negative quadratic correction, in agreement with the behavior found in \cite{Bina:2010ir}.

Equations \eqref{eq:Mn2} and \eqref{eq:nmax} show that GUP quantization of the BH in the reduced phase space produces \emph{a finite spectrum} with an upper bound on the mass of the BH,
\begin{equation}
 M_{\max}^{2}<\frac{m_{\rm P}^{2}}{2\beta},
 \label{eq:Mmax-bound}
\end{equation}
which is never saturated by any discrete level. The existence of a maximum mass arises directly from the maximal momentum encoded in \eqref{eq:GUP}.
%%%%%%%%%%%%%%%%%%%%%%%%%%%%%%%%%%%%%%%%%%%%%%%%%%%%%%%%%%%%%%%%%%%%%%%
\subsection{Relation to polymer quantization and LQG area spectra}

The appearance of a finite and discrete set of admissible masses connects naturally to
other approaches in which quantum geometry introduces an intrinsic ultraviolet scale.  
In polymer quantization \cite{Ashtekar:2002sn, Corichi:2007tf}, the momentum operator is
replaced by a periodic function, $p \rightarrow \frac{\sin(\lambda p)}{\lambda}$, which
imposes a maximal momentum and compactifies the effective phase space.  
This structure is mathematically analogous to the higher-order GUP algebra used here,
$[\hat x,\hat p]= i/(1-\beta \hat p^{2})$, where the commutator diverges at
$|p|=1/\sqrt{\beta}$, forcing the momentum domain to be compact.  
In both frameworks, the compactified momentum space truncates the oscillator spectrum
and produces a finite set of eigenstates, yielding a finite-dimensional Hilbert space
for a fixed macroscopic parameter.

A similar finiteness arises in LQG, in which geometric operators such as area and volume exhibit discrete spectra
\cite{Rovelli:1994ge,Ashtekar:1996eg}.  
In particular, the area operator has eigenvalues determined by the spins of the punctures that pierce the horizon, resulting in a discrete and countable set of horizon areas.  
BH entropy in LQG is obtained by counting these area eigenstates
\cite{Ashtekar:1997yu, Domagala:2004jt, Meissner:2004ju}, and only finitely many microstates exist for a fixed classical area.  
This parallels the finite sequence of mass eigenvalues $M_n$ obtained in our GUP-deformed Schwarzschild model: while the underlying microscopic degrees of freedom differ significantly between GUP and LQG, both frameworks imply that a BH with fixed macroscopic characteristics occupies a finite-dimensional Hilbert space.  
Such finiteness has important implications for unitarity, information recovery, and the absence of infinite-entropy remnants.  
In our case, the truncation of the spectrum arises solely from the GUP-induced maximal momentum and requires no additional assumptions about the microscopic structure of spacetime.

%%%%%%%%%%%%%%%%%%%%%%%%%%%%%%%%%%%%%%%%%%%%%%%%%%%%%%%%%%%%%%%%%%%%%%%%%%%%%%%%%%%%%%%%%%%
\subsection{Physical implications of a finite BH Hilbert space}

A central outcome of the GUP-deformed mass spectrum is that the allowed values of the mass terminate at a finite $n_{\max}$, implying a finite number of microstates and hence a finite-dimensional Hilbert space for the Schwarzschild BH. This feature is highly nontrivial and places the present construction in close conceptual proximity to other quantum-gravitational frameworks in which geometric operators possess discrete and bounded spectra. For example, loop quantum gravity predicts a discrete area operator with a minimum nonzero eigenvalue, producing an effective cutoff in the number of admissible horizon configurations. The GUP-induced mass spectrum found here plays an analogous role: the combination of minimal length and maximal momentum quantizes $M^2$ and prohibits arbitrarily large horizon areas.

A finite Hilbert space also has immediate implications for unitarity. Since the BH cannot evaporate beyond its ground state and cannot grow without bound, all quantum transitions occur within a compact state space. This removes the infinite tower of trans-Planckian modes that typically complicates discussions of BH information loss. In the present framework, the endpoint of evaporation corresponds not to a long-lived or metastable remnant but to the lowest allowed level of the finite spectrum, thereby avoiding the standard challenges associated with remnants carrying arbitrarily large information capacity.

The upper mass bound $M_{\max}$ emerging from the maximal momentum of the GUP algebra can be interpreted as an infrared cutoff for Schwarzschild geometries. In ordinary GR, the Schwarzschild family extends to arbitrarily large $M$, but in the GUP-deformed theory, BHs occupy only the interval $M_{\min}\le M \le M_{\max}$. This is conceptually similar to infrared truncations that appear in compact quantum systems or in approaches to quantum gravity, where the physical Hilbert space is finite. In particular, the existence of $M_{\max}$ implies that the gravitational degrees of freedom effectively ``saturate’’ beyond a critical curvature scale: adding energy no longer increases the size of the BH but instead drives the system toward the terminal level of the spectrum.

Taken together, these features demonstrate that the GUP-quantized Schwarzschild BH behaves as a genuinely finite quantum system with a bounded number of accessible states. This provides a comprehensive model, encompassing ultraviolet and infrared aspects, in which entropy, mass, and temperature are all regulated by the same underlying quantum-gravitational mechanism.

%%%%%%%%%%%%%%%%%%%%%%%%%%%%%%%%%%%%%%%%%%%%%%%%%%%%%%%%%%%%%%%%%%%%%%%
\section{Entropy, temperature, and thermodynamic stability}\label{Sec3}

The adiabatic invariant $I=\oint p\,\mathrm dx$ of the reduced oscillator is quantized semiclassically as $I(M_n)\approx 2\pi(n+\tfrac12)$ \cite{Kunstatter:2002pj, Jalalzadeh:2025uuv}. Identifying the entropy with this invariant (up to a constant) and enforcing consistency with the Bekenstein--Hawking area law in the $\beta\ll1$ limit leads to the GUP entropy
\begin{equation}
 S_{\rm GUP}(M)
 = \frac{4\pi M^{2}}{m_{\rm P}^{2}}
   - \frac{4\pi\beta M^{4}}{m_{\rm P}^{4}}.
 \label{eq:SGUP}
\end{equation}
The first term is the usual area law; the second is a negative quartic correction that dominates at large $M$.

The temperature follows from the first law of thermodynamics, $\mathrm{d}S=\mathrm{d}M/T$:
\begin{equation}
 T_{\rm GUP}(M)
 = \left(\frac{\mathrm{d}S_{\rm GUP}}{\mathrm{d}M}\right)^{-1}
 = \frac{m_P^2}{8\pi M}\,
   \frac{1}{1-2\beta M^{2}/m_{\rm P}^{2}}.
 \label{eq:TGUP}
\end{equation}
For $M^{2}\ll m_{\rm P}^{2}/(2\beta)$, one recovers $T_{\rm H}=m_{\rm P}^{2}/(8\pi M)$; at larger masses, the GUP factor enhances the temperature.

\begin{figure}[t]
\centering
\begin{tikzpicture}
\begin{axis}[
 width=\columnwidth,
 height=0.7\columnwidth,
 xlabel={$M/m_P$},
 ylabel={$T\,/\,m_P$},
 xmin=0.1, xmax=4,
 ymin=0, ymax=0.6,
 domain=0.1:4.9,    % <--- avoid the pole at x=5
 samples=200,
 legend style={at={(0.97,0.97)},anchor=north east},
 thick
]
% standard Hawking
\addplot[solid] {1/(8*pi*x)};
\addlegendentry{$T_H = \frac{m_P^2}{8\pi M}$};

% GUP temperature with beta = 0.02 (illustrative)
\addplot[dashed] {1/(8*pi*x) / (1 - 0.06*x^2)};
\addlegendentry{$T_{\mathrm{GUP}}$ (illustrative $\beta$)};
\end{axis}
\end{tikzpicture}
\caption{Illustrative comparison between the standard Hawking temperature $T_H$ and the GUP-modified temperature $T_{\rm GUP}$ for a representative value of $\beta$ (here $\beta=0.03$ in natural units for visualization only). In the physical regime relevant to astrophysical BHs, $\beta$ is extremely small, and the deviation becomes significant only near the upper end of the spectrum.}
\label{fig:T}
\end{figure}
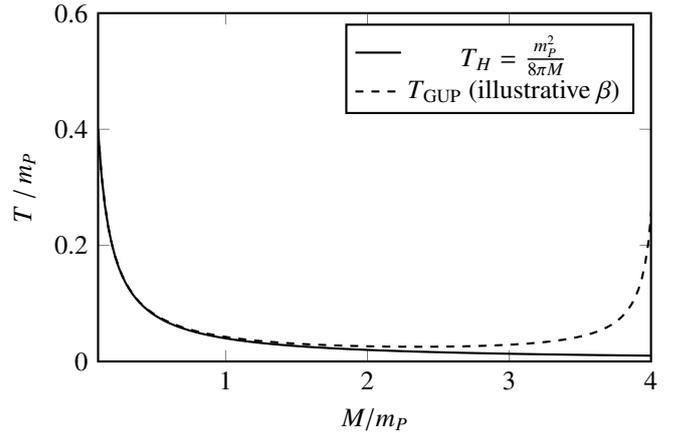

Figure~\ref{fig:T} illustrates $T_{\rm GUP}(M)$ for a representative value of $\beta$ compared to the standard Hawking temperature $T_H\propto 1/M$.

The specific heat,
\begin{equation}
 C_{\rm GUP}(M)
 = \frac{\mathrm{d}M}{\mathrm{d}T_{\rm GUP}}
 = -\,\frac{8\pi\bigl[M(1-2\beta M^{2}/m_{\rm\rm P}^{2})\bigr]^{2}}
         {m_{\rm P}^{2}\bigl(1-6\beta M^{2}/m_{\rm P}^{2}\bigr)},
 \label{eq:C}
\end{equation}
diverges and changes sign at
\begin{equation}
 M_{c}^{2}=\frac{m_{\rm P}^{2}}{6\beta},
 \qquad
 T_{\min}=T_{\rm GUP}(M_{c})
 = \frac{3m_P}{16\pi}\sqrt{6\beta}.
 \label{eq:Mc}
\end{equation}

The GUP-corrected heat capacity, $C_{\rm GUP}(M)$, exhibits qualitatively new thermodynamic structure absent in the Schwarzschild BH. For small masses, the negative specific heat reproduces the usual instabilities of
Hawking evaporation.  
However, at the critical mass, $M_c = m_{\rm P}/\sqrt{6\beta}$, it diverges and changes sign.
This divergence separates two thermodynamic phases: (i) an unstable Schwarzschild-like branch ($C<0$) and (ii) a stable large mass branch ($C>0$), in which the BH heats up more slowly and remains thermodynamically self-regulating. Such a positive specific heat phase does not appear in semiclassical GR and arises here solely from the GUP-induced compactification of the effective phase space. Notably, the critical point lies strictly below the upper bound $M_{\max}<m_P/\sqrt{2\beta}$ enforced by the discrete mass spectrum, and hence the specific heat remains finite at the true endpoint of evaporation.  
Taken together, these features show that GUP not only regulates the Hawking temperature but also fundamentally reshapes the thermodynamic phase structure of the BH.
%%%%%%%%%%%%%%%%%%%%%%%%%%%%%%%%%%%%%%%%%%%%%

\begin{figure}[t]
\centering
\begin{tikzpicture}
\begin{axis}[
    width=\columnwidth,
    height=0.75\columnwidth,
    xlabel={$M/m_P$},
    ylabel={$C_{\rm GUP}(M)$},
    xmin=0.1, xmax=5,
    ymin=-10, ymax=10,
    samples=400,
    domain=0.1:5,
    thick,
    legend style={at={(0.97,0.03)},anchor=south east},
]
\def\beta{0.03}
\addplot[smooth, ]
{( (x*(1 - 2*\beta*x^2))^2 )
    / ( (1)*(1 - 6*\beta*x^2) )};
\addlegendentry{$C_{\rm GUP}(M)$};
\def\Mc{1/sqrt(6*\beta)}

\addplot[dashed,red] coordinates {(\Mc,-10) (\Mc,10)};
\addlegendentry{$M_c = \frac{m_P}{\sqrt{6\beta}}$};

\end{axis}
\end{tikzpicture}
\caption{Specific heat $C_{\rm GUP}(M)$ for a representative small GUP parameter $\beta=0.03$. The heat capacity diverges at 
$M_c = m_P/\sqrt{6\beta}$, marking the transition between an unstable
Schwarzschild-like branch ($C<0$) and a stable large-mass branch ($C>0$). The divergence approaches the maximal mass $M_{\max}<m_P/\sqrt{2\beta}$ from below,
so the specific heat remains finite at the discrete endpoint of the spectrum. }
\label{fig:CGUP}
\end{figure}
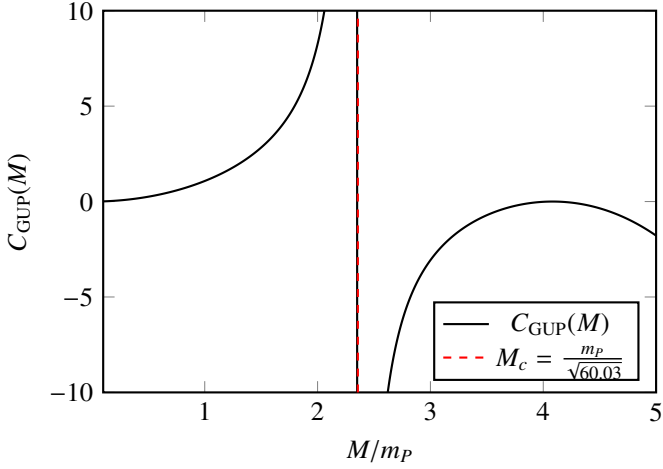

%%%%%%%%%%%%%%%%%%%%%%%%%%%%%%%%%

Finally, since $M^{2}$ never reaches $m_{\rm P}^{2}/(2\beta)$, Eq.~\eqref{eq:SGUP} implies a maximal entropy,
\begin{equation}
 S_{\max}=S_{\rm GUP}(M_{\max})<\frac{\pi}{\beta},
\end{equation}
so the BH Hilbert space dimension $\dim\mathcal{H}_{\rm BH}\sim e^{S_{\max}}$ is finite. The GUP-deformed Schwarzschild BH is thus a finite quantum system: it has finitely many mass levels and microstates.

%%%%%%%%%%%%%%%%%%%%%%%%%%%%%%%%%%%%%%%%%%%%%%%%%%%%%%%%%%%%%%%%%%%%%%%
\section{GUP-deformed lapse and Newtonian limit}\label{Sec4}

We now embed the thermodynamic results into a deformed spacetime metric. More explicitly, we aim to investigate the impact of the GUP parameter on the geometry of spacetime at the semiclassical level. We adopt the static, spherically symmetric ansatz
\begin{equation}
\mathrm ds^{2}
 = -f(r)\,\mathrm dt^{2}
   + \frac{\mathrm dr^{2}}{f(r)}
   + r^{2}\mathrm d\Omega^{2},
 \label{eq:metric}
\end{equation}
and write
\begin{equation}
 f(r) = \left(1-\frac{2GM}{r}\right)\bigl[1+H(r)\bigr],
 \label{eq:fH}
\end{equation}
where $H(r)$ encodes GUP corrections. We impose two conditions: (i) the ADM mass remains $M$, and (ii) the horizon radius is fixed at $r_h=2GM$.

The surface gravity is $\kappa=\frac{1}{2}\mathrm{d}f(r)/\mathrm{d}r|_{r_h}$, and the Hawking temperature is $T=\kappa/(2\pi)$. A short calculation gives
\begin{equation}
\frac{\mathrm{d}f(r)}{\mathrm{d}r}\Bigg|_{r_h} = \frac{1+H(r_h)}{2GM}.
\end{equation}
Matching this to $T_{\rm GUP}(M)$ in \eqref{eq:TGUP} yields
\begin{equation}
  H(r_h)
 = \frac{2\beta M^{2}/m_{\rm P}^{2}}
        {1-2\beta M^{2}/m_{\rm P}^{2}}.
\end{equation}
Requiring $H(r\to\infty)\to 0$ faster than $1/r$ to preserve the ADM mass leads to the minimal profile
\begin{equation}
\begin{split}
 H(r) &= H(r_h)\left(\frac{2GM}{r}\right)^{2}\\
& = \frac{2\beta M^{2}/m_{\rm P}^{2}}
        {1-2\beta M^{2}/m_{\rm P}^{2}}
   \left(\frac{2GM}{r}\right)^{2}.
 \label{eq:H}
 \end{split}
\end{equation}
Substituting into \eqref{eq:fH}, we obtain the GUP-deformed lapse
\begin{equation}
 f(r)
 =
 \left(1-\frac{2GM}{r}\right)
 \left[
 1+
 \frac{2\beta M^{2}/m_{\rm P}^{2}}
      {1-2\beta M^{2}/m_{\rm P}^{2}}
 \left(\frac{2GM}{r}\right)^{2}
 \right].
 \label{eq:fGUP}
\end{equation}
As $r\to\infty$, the geometry is asymptotically flat, near the horizon, the redshift structure is significantly modified so that the geometric temperature matches the GUP result.

In the weak field limit, $g_{tt}=-f(r)\simeq-(1+2\Phi)$, so $\Phi(r)=(f(r)-1)/2$. Expanding \eqref{eq:fGUP} for $GM/r\ll 1$ gives
\begin{multline}
 \Phi_{\rm GUP}(r)
 = -\frac{GM}{r} \\
 + 4\beta\left(\frac{M}{m_P}\right)^{2}
   \left(\frac{GM}{r}\right)^{2}
   \left(1-\frac{2GM}{r}\right)
 + \mathcal{O}\!\left(\frac{1}{r^{4}}\right).
 \label{eq:Phi}
\end{multline}
The leading term reproduces Newton's law; GUP contributes only short-range corrections scaling as $(GM/r)^2$ and higher powers, fully negligible for realistic $\beta$.

%%%%%%%%%%%%%%%%%%%%%%%%%%%%%%%%%%%%%%%%%%%%%%%%%%%%%%%%%%%%%%%%%%%%%%%

\subsection{Astrophysical constraints on the GUP parameter}

The upper bound on the BH mass implied by the GUP spectrum can be confronted with observations of the most massive known SMBHs. In practice, SMBH masses are inferred through several complementary techniques. For nearby galaxies, stellar and gas dynamical modeling of high-resolution kinematic data provides robust estimates based on the gravitational influence of the BH on its environment. At higher redshifts, where spatially resolved dynamics are not available, virial methods applied to broad emission lines in active galactic nuclei are commonly used. In these cases, the BLR size is estimated from either reverberation mapping or empirical radius–luminosity relations, and the mass follows from $M_{\rm vir} \sim f\,R_{\rm BLR} \Delta V^{2}/G$, where $\Delta V$ is the line width and $f$ is a geometrical virial factor calibrated from local AGN samples.

For extremely massive quasars such as TON~618 and SMSS~J2157$-$3602, current mass estimates rely on these virial techniques. Near-infrared spectroscopy of TON~618 suggests a BH mass of order $(4$--$7)\times 10^{10} M_{\odot}$ from its broad emission lines \cite{Shemmer:2004ph}, while SMSS~J2157$-$3602 is inferred to host a BH with $M \simeq (3.4 \pm 0.6)\times 10^{10} M_{\odot}$ \cite{Onken2020}. These values carry systematic uncertainties from the choice of line, the virial factor, and the scatter in the radius-luminosity relation, typically at the level of $\sim 0.3$ dex. Nevertheless, they provide conservative lower bounds on the true maximum BH mass currently observed in the universe.

In our framework, the GUP-imposed upper bound $M_{\max}$ must satisfy $M_{\max} \ge M_{\rm obs}$, where $M_{\rm obs}$ is the mass of the most massive reliably measured SMBH. Using the formal continuum bound $M_{\max}^{2} < m_{\rm P}^{2}/(2\beta)$, this requirement leads to $\beta \lesssim \frac{m_{\rm P}^{2}}{2 M_{\rm obs}^{2}}$.
Taking $M_{\rm obs} \sim 5\times 10^{10} M_{\odot}$ as a representative value yields
\begin{equation}
    \beta \lesssim \mathcal{O}(10^{-98}).
\end{equation}

The systematic uncertainties in the SMBH mass scale translate into at most order-of-magnitude variations in this bound, which does not affect the conclusion that the GUP deformation must be extremely small. Future discoveries of even more massive quasars, or more precise dynamical measurements for existing candidates, would immediately tighten this constraint by lowering the allowed value of $\beta$. In this sense, the upper mass bound predicted by the GUP can be viewed as a phenomenological target for high-mass SMBH surveys and a potential bridge between minimal-length quantum gravity and observational astrophysics.

%%%%%%%%%%%%%%%%%%%%%%%%%%%%%%%%%%%%%%%%%%%%%%%%%%%%%%%%%%%%%%%%%%%%%%%%
\subsection{Newtonian limit and connection with quantum-corrected gravity}

The leading correction to the metric component in \eqref{eq:Phi} scales as $(GM/r)^2$.  
This structure mirrors the universal long-distance quantum-gravity corrections obtained in effective field theory (EFT).  
In particular, the one-loop graviton-exchange calculation yields the modified Newtonian potential
\begin{equation}
\Phi_{\rm EFT}(r)
= -\frac{GM}{r}\left[1 + \alpha\,\frac{GM}{r} + \cdots\right],
\end{equation}
where $\alpha$ is a calculable, theory-independent coefficient \cite{Donoghue1994,BjerrumBohr2003,Khriplovich2002}.  
The GUP correction has the same functional dependence, differing only in the magnitude of the coefficient, which is suppressed by  
$\beta(M/m_P)^2$.  
Thus, the GUP-deformed metric is fully compatible with the EFT structure of quantum gravitational corrections at large distances.

A second close analogy arises in brane world gravity.  
In the Randall--Sundrum scenario, the effective $4D$ exterior metric takes a Reissner--Nordstr\"om form with a \emph{negative} tidal charge $Q_{\rm tid}$ \cite{Dadhich2000}.  
The corresponding potential contains a term $\propto Q_{\rm tid}/r^{2}$, modifying the redshift structure near the horizon while preserving the leading Schwarzschild behavior at large radii.  
The GUP correction in Eq.~\eqref{eq:Phi} plays an analogous role: the effective $(GM/r)^{2}$ contribution behaves like a small induced tidal charge whose magnitude is fixed by the quantum parameter $\beta$.

Solar system experiments tightly constrain deviations proportional to $1/r$ but allow corrections of order $(GM/r)^{2}$ provided their coefficients are sufficiently small.  
Using the astrophysical constraint $\beta \lesssim 10^{-98}$ derived in this work, the GUP-induced modifications remain many orders of magnitude below current observational bounds, even for the most compact stellar-mass and supermassive BHs.  
Hence, the GUP-deformed lapse function is consistent with all weak-field tests while providing a distinctive near-horizon signature rooted in minimal-length quantum gravity.

%%%%%%%%%%%%%%%%%%%%%%%%%%%%%%%%%%%%%%%%%%%%%%%%%%%%%%%%%%%%%%%%%%%%%%%
\section{Conclusions}\label{Sec5}

By quantizing the reduced phase space of the Schwarzschild BH using a higher-order generalized uncertainty principle, we have demonstrated that minimal length quantum gravity can be implemented in a fully canonical and self-consistent manner.  The entire structure follows directly from deforming the fundamental commutator on the BH’s physical phase space, without invoking heuristic arguments or semiclassical ansatze.  This construction yields an array of striking consequences.

First, the spectrum of admissible BH masses becomes both discrete and finite.  The GUP induces a maximal momentum, compactifying the momentum
domain and truncating the oscillator spectrum obtained from the canonical transformation of the Schwarzschild geometry.  As a result, the BH possesses a strict upper bound on the mass, $M_{\max}^{2}<m_{P}^{2}/(2\beta)$, and a nonzero minimum mass.  Because the adiabatic invariant coincides with the BH entropy, the spectrum termination
implies a bounded entropy and therefore a finite-dimensional Hilbert space.  This places the present framework in conceptual alignment with
other quantum geometric approaches (such as polymer quantization and loop quantum gravity), where geometric operators admit discrete spectra and the horizon
area occupies only a finite set of eigenvalues below a fixed bound.

Second, the GUP-regulated temperature derived from the first law remains finite at the endpoint of evaporation.  The specific heat develops a divergence and sign change at a critical mass $M_{c}$, introducing a thermodynamically
stable branch that is absent in the classical Schwarzschild case.  This novel phase structure arises solely from the compactified phase space and does not require exotic matter or modified Einstein equations. The absence of a
divergent temperature or infinite entropy near the endpoint provides a natural resolution of the problematic infinite information remnant scenario.

Third, we constructed a GUP-deformed lapse function $f(r)$ directly from the requirement that the surface gravity reproduce the GUP temperature while preserving the ADM mass and horizon location.  The resulting metric represents
a minimal and internally consistent geometric encoding of GUP physics.  Its weak field limit yields corrections proportional to $(GM/r)^{2}$, fully compatible with current solar system and astrophysical tests, and qualitatively
similar to known quantum-corrected Schwarzschild potentials (e.g., in EFT or brane world tidal charge scenarios).  These short-range corrections leave long-distance phenomenology untouched while modifying the
near-horizon redshift structure in a controlled manner.

Finally, by confronting the theoretical upper mass $M_{\max}$ with observed masses of the most massive SMBHs, we obtained the bound $\beta\lesssim 10^{-98}$.  This demonstrates that astrophysical data (normally
considered far from the Planck scale) can already place highly nontrivial constraints on minimal length quantum gravity.

Altogether, this work provides a unified picture in which the GUP, when applied consistently to the reduced phase space of BHs, simultaneously predicts a finite quantum spectrum, bounded entropy, regulated temperature, and a
geometrically well-defined metric deformation.  The framework is readily extendable to charged or rotating BHs and to cosmological settings, potentially opening a bridge between microscopic quantum gravity models and macroscopic
gravitational phenomenology.

%

%%%%%%%%%%%%%%%%%%%%%%%%%%%%%%%%%%%%%%%%%%%%%%%%%%%%%%%%%%%%%%%%%%%%%%%%%%%%%%%%%%%%%%%%%%%%%%%%%%%%%%%%%%%%
\section*{Data availability}
No data was used for the research described in the article.

\section*{Declaration of competing interest}
The authors declare that they have no known competing financial interests or personal relationships that could have appeared to influence the work reported in this paper.

%\appendix

%\section{Appendix title 1}

%\section{Appendix title 2}

\bibliographystyle{elsarticle-num} 
\bibliography{GUP-Black}



%% For example:

%% \bibitem[Aladro et al.(2015)]{Aladro15} Aladro, R., Martín, S., Riquelme, D., et al. 2015, \aas, 579, A101

%%\end{thebibliography}

\end{document}